\documentclass[article]{aa}
\usepackage{enumitem}
\usepackage{graphicx}
\usepackage{cancel}
\usepackage{txfonts}
\usepackage[pdfencoding=auto,psdextra]{hyperref}
\usepackage{pifont}
\usepackage[export]{adjustbox}
\usepackage[dvipsnames]{xcolor}

\usepackage[dvipsnames]{xcolor}
\hypersetup{
    colorlinks=true,
    linkcolor=BrickRed,
    anchorcolor=BrickRed,
    filecolor=Violet,      
    urlcolor=Mulberry,
    citecolor=MidnightBlue,
}
\usepackage{txfonts}
\usepackage{lipsum}
\usepackage{subcaption}         
\usepackage{lscape} 
\usepackage{placeins}           

\begin{document}
\makeatletter
\let\linenumbers\relax
\let\pagewiselinenumbers\relax
\@ifundefined{nolinenumbers}{}{\nolinenumbers}
\makeatother
   \title{Two-dimensional galaxy cluster shapes using lensing data}

   \subtitle{A simulated analysis}

   \author{C. Giocoli\inst{1,2}\corrauth{carlo.giocoli@inaf.it}        
       \and M. Meneghetti\inst{1,2} \and G. Despali\inst{3,1,2}  \and L. Moscardini\inst{3,1,2} \and E. Rasia\inst{4,5} \and G. Lesci\inst{6} \and W. Cui\inst{7,8} \and G. Yepes\inst{7,8}}

   \institute{
     INAF-Osservatorio di Astrofisica e Scienza dello Spazio di Bologna, Via Piero Gobetti 93/3, I-40129 Bologna, Italy
     \and
     INFN -- Sezione di Bologna, viale Berti Pichat 6/2, I-40127 Bologna, Italy
     \and
     Dipartimento di Fisica e Astronomia "Augusto Righi", Alma Mater Studiorum Università di Bologna, via Gobetti 93/2, I-40129 Bologna, Italy  
     \and 
     Department of Physics; University of Michigan, Ann Arbor, MI 48109, USA
     \and 
     INAF -- Osservatorio Astronomico di Trieste, via Tiepolo 11, I-34131, Trieste, Italy
     \and 
     INAF -- Istituto di Radioastronomia, Via Piero Gobetti 101, I-40129, Bologna, Italy
     \and
    Departamento de Física Teórica, Módulo 15, Facultad de Ciencias, Universidad Autónoma de Madrid, E-28049 Madrid, Spain 
    \and 
    Centro de Investigación Avanzada en Física Fundamental (CIAFF), Facultad de Ciencias, Universidad Autónoma de Madrid, E-28049 Madrid, Spain     
    }

   \date{Received: {\textbf{\color{magenta} Today}}, Published: {\textbf{\color{cyan}Tomorrow}}.\, \emph{\href{https://www.youtube.com/watch?v=ap0IB0F4Jak}{After all... tomorrow is another day}}}
 
  \abstract
   {The morphological properties of galaxy clusters reflect their assembly history and dynamical state.   High-resolution weak lensing data allow us to reconstruct individual structural and morphological properties, highlighting departure from spherical symmetry. While weak lensing is typically used to recover cluster masses, its ability to recover the two-dimensional shape is less explored. }
   {We investigate the reliability of weak lensing in recovering the projected cluster morphologies and study how the shape depends on the particular cluster orientation and dynamical state. In addition, we compare the morphology probed by weak lensing -- 
    on larger scales, with the one traced by strong lensing -- towards the cluster centre.} 
   {We study the structural shape parameters of three different clusters at $z$=0.22 from the Three Hundred Project selected to represent a relaxed, mildly disturbed and an unrelaxed system. For each cluster, we consider three random projections and three other projections where the system is oriented along the principal axes of the mass tensor. For each projection, we generate 128 independent weak lensing catalogues assuming a source density of 30 arcmin$^2$ and an intrinsic ellipticity dispersion of $\sigma_{\epsilon}=0.3$. We first constrain the mass and concentration from the one-dimensional lensing signal using NFW and BMO profiles, and subsequently recover the projected ellipticity, position angle, and centre using a two-dimensional elliptical NFW lens model. In addition, we measure the morphology from the true convergence maps, from  Kaiser \& Squires convergence reconstructions, and from the tangential critical lines in the strong-lensing region.}
   {We find that the recovered projected parameters depend on the cluster orientation and dynamical state. While for the relaxed cluster the inner and large-scale shapes are relatively similar, for the two disturbed systems, their central region appears more elliptical. The two-dimensional weak-lensing modelling recovers the orientation of the projected mass distribution despite the scatter introduced by source sampling and intrinsic shape noise. The Kaiser \& Squires reconstruction recovers the cluster position angle quite well while preferring a rounder projected mass density distribution. Larger centroid offsets are found for the unrelaxed systems.}
   {Combining weak-lensing morphology with the inner structure traced by strong lensing provides a complementary view of cluster structure across physical scales and offers a promising outlook towards using lensing as a probe of cluster morphology and dynamical state.}


   \maketitle

\section{Introduction}

Galaxy clusters are the most massive and late-forming gravitationally bound structures in our Universe.
Their hierarchical assembly allows us to trace the growth of cosmic structures and highlights the physical 
processes governing their formation \citep{white79,white93,white96b,tormen98a,borgani11}. Dark matter haloes form as a consequence of repeated merging events 
of small systems, preferentially occurring along the filamentary structure of the cosmic web \citep{bond91,kauffmann93,baugh06,fakhouri10}. 
This highlights that dark matter haloes are not expected to be spherical: their three-dimensional mass 
distributions are generally triaxial \citep{maccio07,munozcuartas11,schneider12b,vega-ferrero17}, and their shape reflects their 
formation histories, dynamical state and the surrounding correlated large-scale structure environment \citep{maccio08}. 

The departure from spherical symmetry in dark matter haloes is rooted in the theory of structure formation. The statistics of Gaussian random fields provide the theoretical model for predicting the geometry of the density fluctuations from which dark matter haloes form \citep{doroshkevich70,bardeen86,bond96}. 
\citet{sheth01b} and \citet{sheth02}  integrated the ellipsoidal collapse model in the excursion-set formalism. In this context, the evolution of a perturbation depends not only on its initial overdensity but also on the ellipticity and prolateness associated with the eigenvalues of the deformation tensor \citep{rossi11}, modifying the threshold for collapse to become mass-dependent, leading to the well-known moving-barrier formulation of the excursion-set model \citep{giocoli07,moreno08,moreno09}.

Cosmological numerical simulations have established the
triaxial nature of dark-matter haloes and its connection
to hierarchical structure formation
\citep{jing02,despali13}.
In particular, more massive and higher-redshift haloes
tend to be less spherical, with their shapes showing
a nearly universal dependence on the peak-height parameter
\citep{despali14,bonamigo15}.
Subsequent studies have further characterised halo
shapes and their evolution
\citep{despali17,vega-ferrero17,okabe20a,okabe20b}.

Observationally, gravitational lensing provides a direct
probe of the projected mass distribution of galaxy clusters
\citep{bartelmann01,bartelmann10}.
Although weak lensing is widely used to measure cluster
masses and calibrate mass--observable relations, most
analyses rely on azimuthally averaged tangential shear
profiles, which erase information about the projected
halo morphology. This information can instead be
extracted from the full two-dimensional shear field,
through direct modelling of individual clusters or
multipole and stacked analyses
\citep{umetsu20a,zhang23,okabe25,okabe26}.

One of the most direct demonstrations of this approach was presented by \citet{oguri10}
who studied the two-dimensional weak-lensing shear map of 25 massive clusters observed with Subaru. By fitting elliptical halo models directly to the measured two-dimensional shear pattern, they simultaneously constrained cluster mass, concentration, centroid, ellipticity, and position angle. 
They obtained a significant detection of the ellipticity of the dark-matter distribution and found its distribution to be broadly consistent with expectations for collisionless CDM haloes. 

Numerical studies have long shown that halo triaxiality and structures along the line of sight substantially contribute to the scatter of weak-lensing mass estimates \citep{bahe12,giocoli24,ingoglia24,sommer24,giocoli25,lesci26}
In addition, it is worth underlining that by projecting clusters along the principal axes of their matter distributions, the orientation modulates both the lensing signal and the recovered mass: projections along the halo major axis enhance the weak-lensing signal and tend to produce larger inferred masses. This result further motivates moving beyond the conventional azimuthally averaged lensing analysis. 

In this work, we investigate the recovery of galaxy cluster
morphology from weak lensing and its dependence on dynamical
state and projection effects. We analyse three clusters
from The Three Hundred Project \citep{cui18}, spanning
relaxed, mildly unrelaxed, and strongly unrelaxed dynamical
states. For each cluster, we generate strong- and weak-lensing
maps along six lines of sight: three random orientations and
three aligned with the principal axes of the three-dimensional
mass tensor. Comparing the weak- and strong-lensing
measurements allows us to investigate cluster morphology
across different physical scales. This controlled sample is
not statistically representative, but enables us to isolate
projection effects and assess the influence of dynamical
state on the recovery of projected cluster shapes.

The paper is organised as follows. In Sect.~\ref{Simulation}, we present the simulated clusters used in this work and describe how we construct weak- and strong-lensing simulations. 
In Sect.~\ref{analytical}, we introduce the one- and two-dimensional models and the corresponding likelihoods for describing the weak lensing data. In Sect.~\ref{Results} we present our results and compare the cluster projected morphologies derived from weak and strong lensing data. We summarise and conclude in Sect.~\ref{sandc}. In App.~\ref{app_1} we display the cluster projections along the minor, intermediate and major axis of the mass tensor ellipsoid. In App.~\ref{app_2}, we present two additional two-dimensional posterior corner plots, and in App.~\ref{app_3}, we show a table of the Einstein radii for each cluster projection.

\section{Numerical simulations}
\label{Simulation}

\begin{figure*}
    \centering
    \includegraphics[width=\hsize]{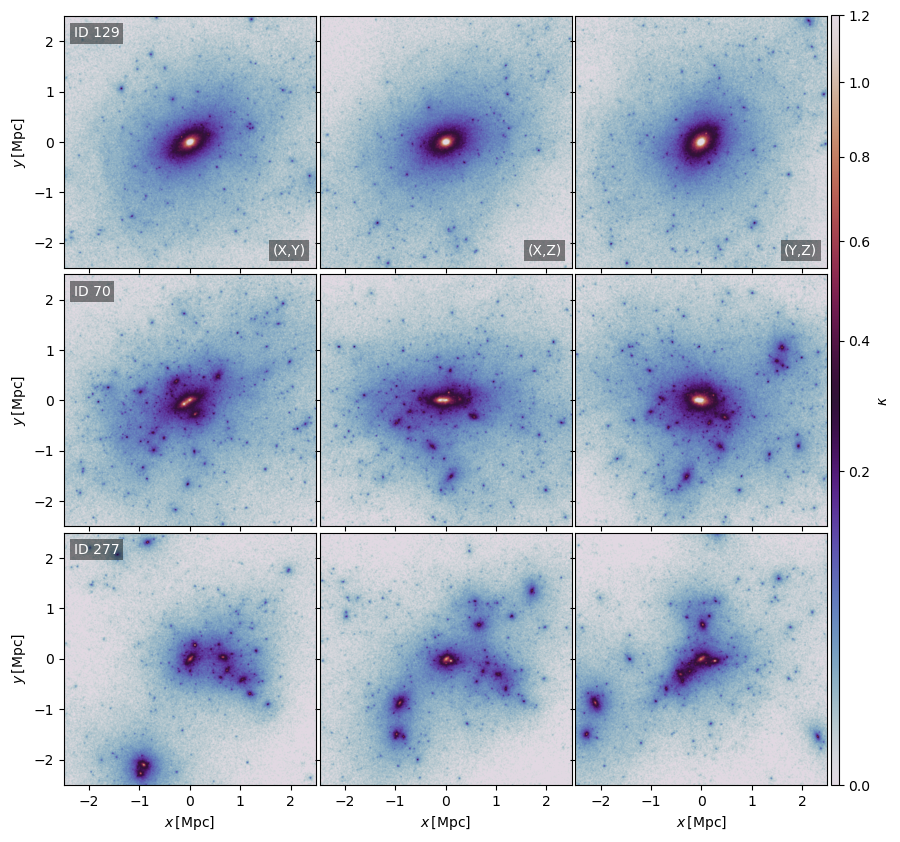}
    \label{Three_Random_Clusters}
    \caption {Convergence maps of the three random projections for the three clusters considered in this work. From top to bottom, we show clusters sorted from the most to the least relaxed. The particles are projected into the lens plane along the simulated box axes: along the $Z$, $Y$ and $X$ axes, from left to right, respectively. Along the line-of-sight, we project the particles present within $\pm$ 5 Mpc of the cluster centre.\label{kappa_random}}
\end{figure*}

\subsection{Three among The Three Hundred Clusters}

The galaxy clusters analysed in this work are part of The Three Hundred Project \citep{cui18}, a suite of hydrodynamical zoom-in simulations specifically designed to study the formation and evolution of massive galaxy clusters and their environments. The sample originates from the MultiDark Planck 2 (MDPL2) dark-matter-only simulation \citep{klypin16}, which follows $3840^3$ dark-matter particles in a periodic box of 1 $h^{-1}{\rm Gpc}$ on a side, assuming a flat $\Lambda$CDM cosmology 
with cosmological parameters consistent with \citet{planck16a}. The 324 most massive haloes identified at $z$=0 were selected, and spherical regions of radius 15 $h^{-1}{\rm Mpc}$ centred on each halo were re-simulated at higher resolution, including baryonic physics. 

In this work, we use the runs performed using \texttt{Gadget-X}, an improved version of the \texttt{GADGET-3} code with an alternative prescription for radiative cooling, star formation, chemical enrichment, feedback from stars and active galactic nuclei and subgrid physics model \citep{rasia15,beck16a}. 

In each re-simulated box, bound structures in haloes and subhaloes are identified using the AMIGA Halo Finder (\texttt{AHF}) \citep{knollmann09}. Cluster masses are defined as:
\begin{equation}
M_{200}=\frac{4\pi}{3}R_{200}^{3}200\,\rho_{\rm c}(z)\,,
\end{equation}
where $R_{200}$ represents the radius at which the enclosed halo density equals $200$ times the critical density $\rho_{\rm c}(z)$.

It is worth mentioning that \texttt{AHF} identifies the halo centre as the local density peak in an adaptively refined manner and then determines the set of gravitationally bound particles associated with the main structure; we therefore adopt the position of the maximum density peak of the halo $\mathbf{R}_{\rm c}=(X_c, Y_c, Z_c)$.

The dynamical state of clusters can be characterised using indicators sensitive to different manifestations of departure from equilibrium \citep{neto07}. Following \citet {cui17,haggar20,herbonnet22}, we consider three complementary quantities: the virial ratio parameter $\eta$, the centre-of-mass offset $\Delta_r$, and the fraction of mass bound to substructures $f_{\rm s}$.
The virial ratio is defined as:
\begin{equation}
    \eta = \frac{2T-E_{\rm s}}{|W|}\,,
\end{equation}
where $T$ and $W$ are the total kinetic and potential energies of the cluster, respectively, and $E_{\rm s}$ accounts for the surface-pressure term associated with material crossing the halo boundary. A system close to virial equilibrium is expected to have $\eta\simeq 1$. 

The centre-of-mass offset can be read as:
\begin{equation}
\Delta_r =
\frac{|\boldsymbol{R}_{\rm cm}-\boldsymbol{R}_{\rm c}|}
{R_{200}}\,,
\end{equation}
where $\boldsymbol{R}_{\rm cm}$ is the centre of mass within $R_{200}$ and $\boldsymbol{R}_{\rm c}$ denotes the cluster centre from \texttt{AHF}. This quantity is particularly sensitive to asymmetric mass distributions and recent merger events: for a relaxed system, the centre of mass and density peak are expected to nearly coincide $\Delta_r \approx 0$, whereas mergers and massive infalling structures can produce a significant displacement between them.

The substructure mass fraction is defined as:
\begin{equation}
    f_{\rm s} =
\frac{\sum_i m_{{\rm sub},i}}{M_{200}}\,,
\end{equation}
measuring the fraction of the total cluster mass within $R_{200}$ contained in gravitationally bound subhaloes, and therefore providing a complementary indication of recent accretion and merger activity.

A cluster is classified as relaxed when the following three conditions, within $R_{200}$, are all met:
\begin{eqnarray}
0.85 < &\eta& < 1.15\,,\\
&\Delta_r& < 0.04\,,\\
&f_{\rm s}& < 0.1\,.
\end{eqnarray}

\begin{table}[ht!]
\caption{\label{TabHaloes}Properties at $z=0.22$ of the clusters considered in this work}
\centering
\begin{tabular}{lcccccc}
\hline\hline 
ID &$M_{200}$ & $c_{200}$&  $1-c/a$ & $\eta$ & $\Delta_r$ & $f_s$   \\
\hline \\
129 & $8.65$ & $6.60$ &0.409 & $0.998$ & $0.017$ & $0.03$ \\
$\;70$ & $10.38$ & $3.48$ & 0.449 & 1.177 & 0.076 & 0.07 \\ 
277 & 4.84 & 3.08 & 0.339& 1.273 & 0.238 & 0.10   \\ \hline
\end{tabular}
\tablefoot{The various rows show the cluster ID, mass $M_{200}$ -- expressed in $10^{14}\;[M_{\odot}/h]$, concentration $c_{200}$\footnote{We compute the concentration by fitting the total density cluster profile using an NFW model. It is worth mentioning that this value is different from the one reported in the \texttt{AHF} catalogue files.}, 3D ellipticity  $1-c/a$ (within $R_{200}$), the virial ratio $\eta$, the centre-of-mass offset $\Delta_r$ and the subhalo mass fraction $f_s$ of the three systems considered in this work.}
\end{table}

For the present analysis, we select three clusters at $z=0.22$ which span different dynamical states. Their main properties are summarised in Table~\ref{TabHaloes}. 
In particular, cluster 129 represents the most dynamically relaxed system of the whole cluster sample. The parameters indicate a regular and relatively undisturbed mass distribution. On the other hand, cluster 277 represents the most dynamically disturbed configuration with a centre-of-mass offset equal to 0.238. Cluster 70 occupies an intermediate regime.  We therefore use clusters 129, 70, and 277 as representative examples of a relaxed, mildly unrelaxed, and strongly unrelaxed system, respectively. This selection allows us to investigate how an increasing degree of dynamical complexity affects the projected cluster morphology and, in particular, the ability of weak lensing to recover the shape of the underlying mass distribution.

For each of the three selected clusters, we consider six projections. Three projections are taken along the Cartesian axes of the simulation volume. Since the orientation of each halo is unrelated to the simulation coordinate system, these provide three orthogonal and effectively random lines of sight. In addition, we construct three particular projections aligned with the principal axes of the cluster mass-tensor distribution: minor, intermediate and major axes of the ellipsoid, as computed by AHF using the halo particle distribution within $R_{200}$.

The combination of random and principal-axis projections is particularly useful for the present study. While the former mimics the arbitrary orientations encountered in observations, the latter provides a controlled experiment in which the relation between the intrinsic three-dimensional geometry of a cluster and its projected morphology is explicitly known. In this latter case, when the cluster tends to be oriented along the minor axis, its projected mass appears more elliptical, while when oriented along the major axis, it seems more circular. Overall, our sample therefore consists of $3\times6=18$ cluster projections, spanning three different dynamical states and a range of viewing orientations.

\subsection{Lensing Simulations}

For each line of sight, the cluster particle distribution is projected onto a two-dimensional lens plane, consistently accounting for contributions of the different matter components to the projected surface mass-density field $\Sigma(\boldsymbol{\theta})$ from which we define the convergence as:
\begin{equation}    
\kappa(\boldsymbol{\theta}) \equiv \frac{\Sigma(\boldsymbol{\theta})}{\Sigma_{\rm crit}}\,,
\end{equation}
where $\Sigma_{\rm crit}$ indicates the critical surface mass density for the adopted lens and source redshifts, which can be read as:
\begin{equation}
  \Sigma_{\rm crit} \equiv \dfrac{c^2}{4 \pi G} \dfrac{D_{\rm s}}{D_{\rm l}
    D_{\rm ls}}\,,
  \label{eq_scrit}
\end{equation}
with $D_{\rm l}$, $D_{\rm s}$, and $D_{\rm ls}$ indicating the observer-lens,
observer-source, and source-lens angular diameter distances,
respectively; $c$ represents the speed of light and $G$ the universal
gravitational constant. Our reference weak lensing maps have been created assuming a fixed source redshift $z_{\rm s}=3$, on a field of view of 5 Mpc on a side and resolved with $2048\times 2048$ pixels. Along the line of sight, maps extend $\pm 5$ Mpc from the cluster centre.

\begin{figure*}[ht!]
\centering
\includegraphics[width=0.74\hsize]{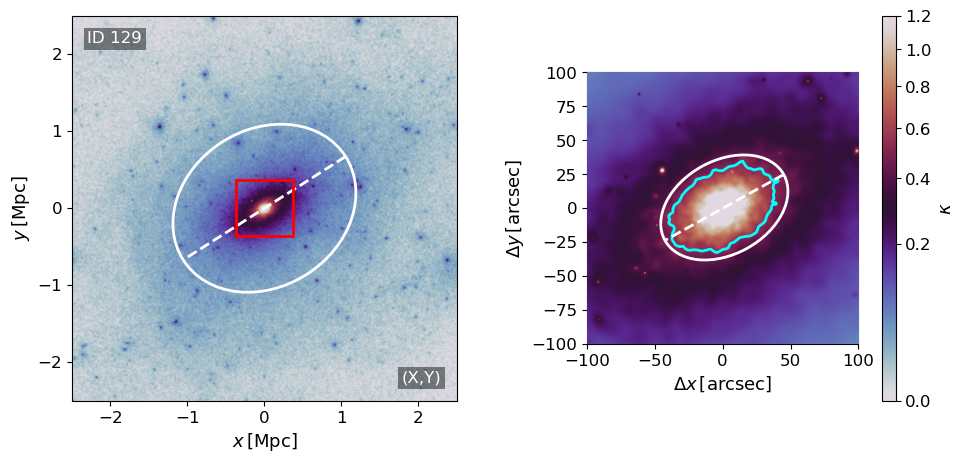}
\includegraphics[width=0.74\hsize]{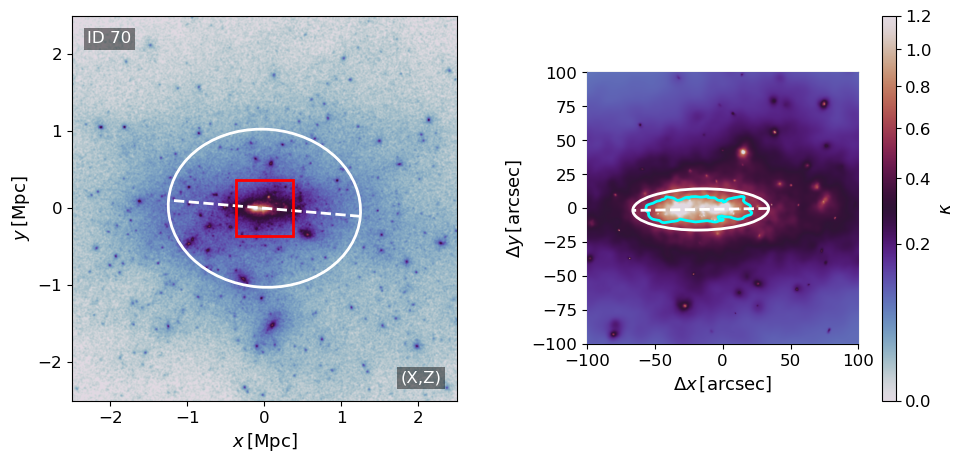}
\includegraphics[width=0.74\hsize]{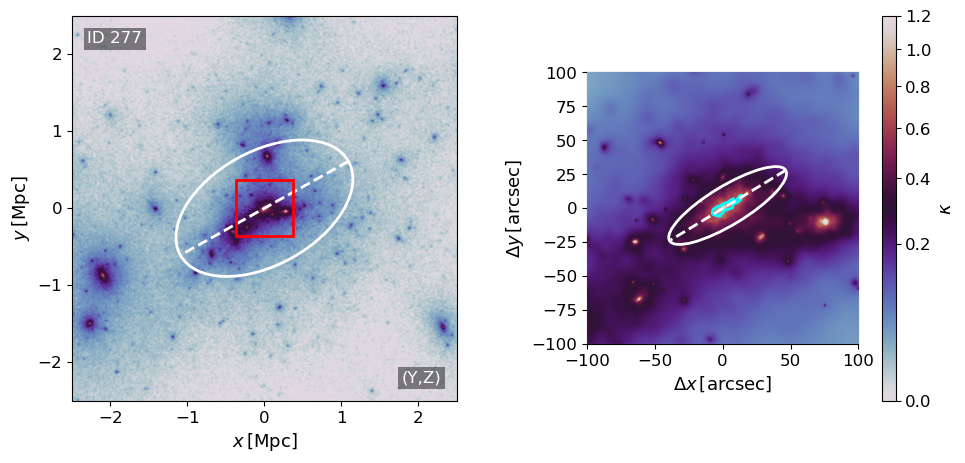}   
\caption{Weak (left) and strong (right) lensing maps for one random projection per cluster. Rows from top to bottom show cluster 129 projection ($X,Y$), cluster 70 projection ($X,Z$) and cluster 277 projection ($Y,Z$), respectively. The ellipses and dashed lines in the left panels display the best-fit morphology of the large-scale projected matter density distribution. The cyan closed curves on the right panels correspond to the central tangential critical lines for $z_s=3$, while the ellipses and the dashed line indicate the best-fit ellipses and position angle to those. For graphical purposes, when plotting the ellipses, we adopted an arbitrary major-axis size of 1.25 Mpc and 50 arcsec for the weak and strong lensing maps, respectively. 
\label{Map_WSL}}
\end{figure*}

In Fig.~\ref{kappa_random} we show the convergence maps of the three random projections (from left to right) going from the relaxed to the unrelaxed one (from top to bottom, respectively). From the figure, we can already see by eye that cluster 129 appears relatively smooth and relaxed, whereas cluster 277 is composed of multiple clumps and is clearly in a merging phase.  
In App.~\ref{app_1} we present the preferential projections of the same clusters.

From the convergence map, we compute the lensing potential $\psi$ using the two-dimensional Poisson equation,
\begin{equation}
\nabla_{\boldsymbol{\theta}}^2
\psi(\boldsymbol{\theta})
= 2\,\kappa(\boldsymbol{\theta}) .
\end{equation}
We solve this equation numerically in Fourier space using a Fast Fourier Transform (FFT). Since the discrete Fourier transform implicitly assumes periodic boundary conditions, discontinuities at the edges of the convergence map may introduce artificial features in the reconstructed lensing fields. To mitigate these boundary effects, we zero-pad the convergence maps by 0.5 $\rm Mpc$ on each side before performing the Fourier-space calculation. The padded region is discarded after the lensing quantities have been computed.
The two components of the shear field $\gamma$ are subsequently obtained from the second derivatives of the lensing potential,
\begin{eqnarray}
\gamma_1(\boldsymbol{\theta})
&=&
\frac{1}{2}
\left[
\frac{\partial^2\psi(\boldsymbol{\theta})}{\partial x^2}
-
\frac{\partial^2\psi(\boldsymbol{\theta})}{\partial y^2}
\right]\,, \\
\gamma_2(\boldsymbol{\theta}) 
&=&
\frac{\partial^2\psi(\boldsymbol{\theta})}
{\partial x \, \partial y}\,,
\end{eqnarray}
where $x$ and $y$ denote the two angular coordinates in the lens plane. 

\begin{figure*}[ht!]
\centering
\includegraphics[width=0.33\hsize]{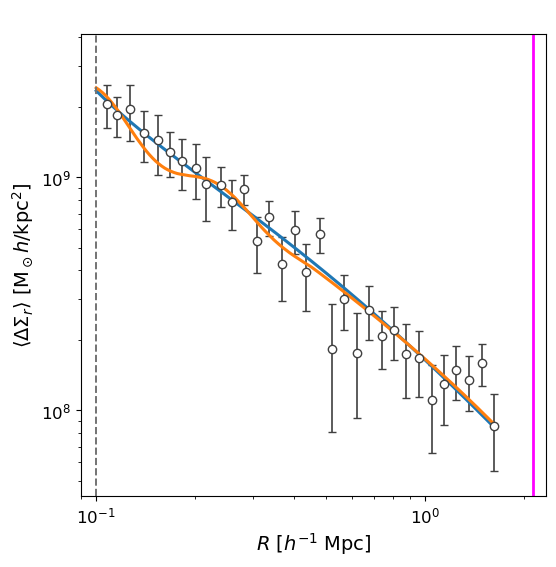}
\includegraphics[width=0.33\hsize]{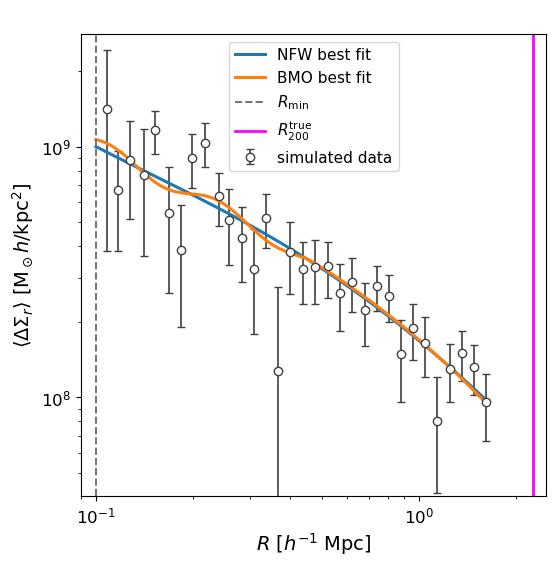}
\includegraphics[width=0.33\hsize]{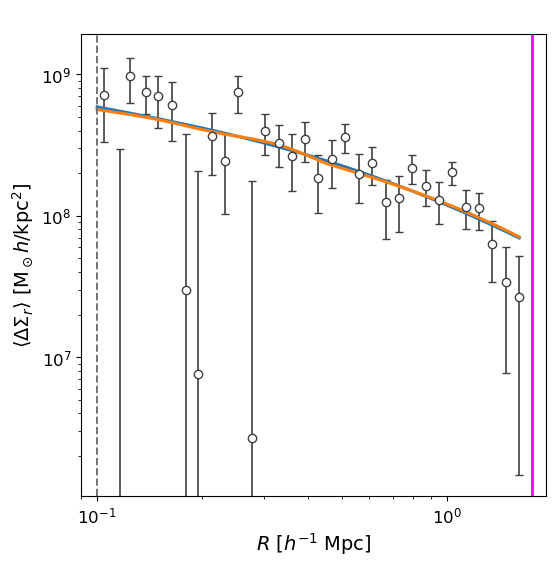}   
\caption{Excess surface mass density profile computed for the same clusters and projections displayed in Fig.~\ref{Map_WSL}. The data points with the error bars show the signal measured from the simulations. Blue and orange curves are the best-fit models adopting the NFW and BMO models, respectively.  
    \label{1D_prof}}
\end{figure*}

In the weak-lensing analysis, we account for the fact that the observable distortion of background galaxy images is related to the reduced shear, rather than directly to the shear. We therefore define the two components of the reduced shear as
\begin{equation}
g_i(\boldsymbol{\theta})
=
\frac{\gamma_i(\boldsymbol{\theta})}
{1-\kappa(\boldsymbol{\theta})},
\qquad i=1,2 .
\end{equation}
In the limit $\kappa\ll1$, the reduced shear approaches the shear, $g\simeq\gamma$, while the distinction becomes increasingly important towards the inner cluster regions. The reduced-shear field is therefore used throughout our weak-lensing pipeline, both in constructing the mock observations and in modelling the lensing signal.
Background sources are randomly distributed across the field of view according to the expected source number density, and the two reduced-shear components, $g_1$ and $g_2$, at their positions are obtained from the simulated maps. We adopt a source population with a number density of $n_{\rm gal}=30\,{\rm arcmin}^{-2}$.
To mimic the statistical uncertainty affecting weak-lensing measurements, we include the contribution from the intrinsic ellipticity of background galaxies by adding an intrinsic shape-noise contribution to the reduced-shear signal with an ellipticity dispersion of $\sigma_{\epsilon}=0.3$. For each cluster projection, we generate 128 independent mock weak-lensing catalogues, each characterised by a different realisation of the background source positions and intrinsic shape noise. Since each cluster is analysed along six different lines of sight, this results in a total of 768 mock shear catalogues per cluster. This ensemble of realisations allows us to quantify the statistical scatter in the recovered weak-lensing properties arising from the finite source density and intrinsic galaxy ellipticities.

We complement the weak lensing maps with corresponding high-resolution strong lensing maps that cover the central $200 \times 200$ arcsec field of view with $2048 \times 2048$ pixels. We use those to compute the principal critical lines, from the condition
\begin{equation}
\lambda_{\rm t}(\boldsymbol{\theta})
=1-\kappa(\boldsymbol{\theta})
-\gamma(\boldsymbol{\theta})=0\,,
\end{equation}
where $\gamma(\boldsymbol{\theta}) = \sqrt{\gamma_1^2(\boldsymbol{\theta}) + \gamma_2^2(\boldsymbol{\theta})}$ -- that contain the cluster centre at three source redshifts $ z_s=1$, 3, and 6. For the analysis presented in this work, we adopt as reference the critical value of the tangential critical lines for source redshift $z_s=3$. In App.~\ref{app_3} we report the values of the Einstein Radius computed for each considered cluster projection. 

As an example, Fig.~\ref{Map_WSL} illustrates the projected mass distribution and the corresponding lensing properties of one projection per cluster. 
The weak-lensing panels (on the left) show the convergence distribution on the scales used for the WL analysis. We characterise the morphology of the underlying projected mass distribution by fitting an ellipse to the convergence map, forcing the centre to be located at $(0,0)$. The resulting ellipse provides our reference estimate of the projected shape of the cluster within the considered field of view, allowing us to quantify its axis ratio and position angle for each line of sight. As expected for a triaxial and structurally complex mass distribution, both quantities vary with the projection considered, even for the same three-dimensional halo.
The central regions, from the strong lensing map, are shown separately on a smaller physical scale on the right of each panel. Here, we display the tangential critical line 
and fit an ellipse. The shape and orientation of this ellipse provide a complementary measure of the projected cluster morphology towards the cluster centre and can be affected by the inner structure, asymmetries, and substructures; to model them, we use the barycentre of the critical lines as the ellipse's centre.
The two ellipses give us an idea of the cluster morphology on two different characteristic scales. Their comparison allows us to investigate whether the orientation and ellipticity of the cluster remain coherent across scales or whether the projected morphology changes from the cluster core to its outskirts. These reference measurements are subsequently compared with the shape recovered by our two-dimensional weak-lensing modelling pipeline, as we will discuss in the next section.  

\begin{figure*}[ht!]
\centering
\includegraphics[width=0.33\hsize]{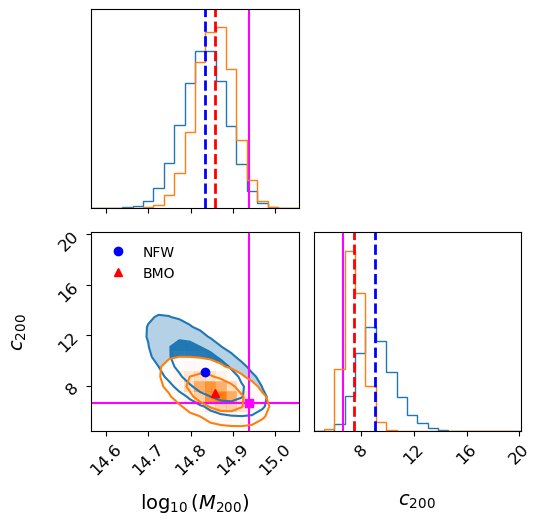}
\includegraphics[width=0.33\hsize]{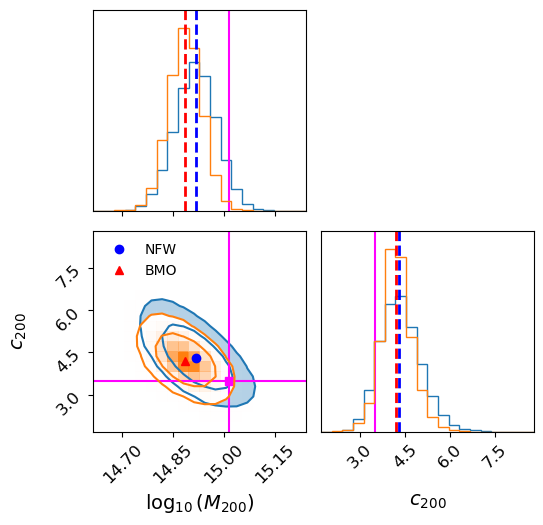}
\includegraphics[width=0.33\hsize]{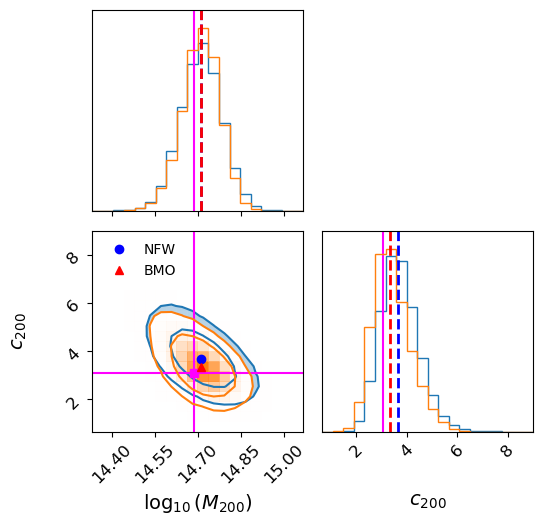}     
\caption{Posterior distributions in $\log(M_{200})$ and $c_{200}$ when fitting the one-dimensional density profiles for the three clusters and the same projections displayed as in Fig.~\ref{1D_prof}. The blue and orange distributions show the results of modelling the excess surface mass density using the NFW and BMO profiles, respectively. The magenta lines indicate the true three-dimensional mass and concentration of the cluster. \label{fig_1DPost}}
\end{figure*}

\section{The analytical models}
\label{analytical}
Before investigating the two-dimensional morphology of the clusters, we first perform a one-dimensional weak-lensing analysis of the averaged lensing signal. This provides a reference measurement of the cluster mass and concentration, allowing us to quantify the impact of the assumed radial density profile on the inferred halo parameters. 
In particular, we compare the results obtained assuming a Navarro--Frenk--White (NFW; \citealt{navarro96,navarro97}) profile with those derived using the smoothly truncated profile proposed by \citet{baltz09}, hereafter BMO. The latter has been adopted in a series of works based on both numerical simulations of clusters \citep{giocoli24,giocoli25} and photometric weak lensing data \citep{giocoli21,lesci22,lesci22b}.
In a second step, we exploit the full two-dimensional reduced-shear field using an elliptical NFW model to constrain the projected cluster morphology, including its ellipticity, position angle, and centroid. It is worth noting that we first perform this analysis using uniform priors on all model parameters, thereby allowing the two-dimensional lensing signal to constrain the cluster's radial and morphological properties simultaneously, without incorporating information from the one-dimensional fit. We then repeat the analysis using Gaussian priors on $\log(M_{200})$ and $c_{200}$, centred on the posterior constraints obtained from the corresponding one-dimensional NFW analysis. This second configuration allows us to investigate how prior knowledge of the radial mass distribution influences the recovery of the projected morphology and, in particular, to assess the degeneracy between mass, concentration, and ellipticity. The comparison between the uniform- and Gaussian-prior analyses therefore provides a direct test of the extent to which the inferred cluster shape is driven by the two-dimensional lensing information itself or is affected by constraints on the underlying radial mass profile.

\subsection{One-dimentional modelling}

For each cluster and line of sight, we compute the tangential component of the reduced shear with respect to the cluster centre and construct the radial lensing profile. The tangential reduced shear can be read as:
\begin{equation}
g_{\rm t}(R) =
-\left[g_1\cos(2\phi)+g_2\sin(2\phi)\right],
\end{equation}
where $\phi$ is the polar angle of each background source with respect to the cluster centre. The signal is then azimuthally averaged in 32 logarithmically spaced radial bins from $R_{\rm min}=0.1\, h^{-1}\mathrm{Mpc}$ up to the cluster half field of view. We conservatively exclude the innermost region from the weak-lensing analysis. At these scales, baryonic processes modify the density profile, and the lensing signal may depart from the reduced-shear weak-lensing regime.
It is also worth mentioning that the high projected density of cluster galaxies in the central regions makes identifying background sources in real observations more difficult due to source blending, complicating the separation of background galaxies from foreground objects and cluster members. 

We adopt two models for the cluster density distribution. The first is the NFW profile \citep{navarro96,navarro97}, which can be read as:
\begin{equation}
\rho_{\rm NFW}(r) =
\frac{\rho_{\rm s}}
{(r/r_{\rm s})(1+r/r_{\rm s})^2},
\end{equation}
where $r_{\rm s}$ and $\rho_{\rm s}$ are the scale radius and characteristic density, respectively. We parametrise the model in terms of $M_{200}$ and the concentration
\begin{equation}
c_{200} = \frac{R_{200}}{r_{\rm s}}.
\end{equation}
The NFW profile provides a simple description of the equilibrium halo density distribution, but its density decreases as $r^{-3}$ at large radii, giving an infinite mass.
As an alternative, we consider the BMO profile \citep{baltz09}, which introduces a smooth truncation of the NFW density distribution,
\begin{equation}
\rho_{\rm BMO}(r)
=
\rho_{\rm NFW}(r)
\left(
\frac{r_{\rm t}^2}
{r^2+r_{\rm t}^2}
\right)^2 ,
\end{equation}
where $r_{\rm t}$ is the truncation radius. We express it as
\begin{equation}
r_{\rm t}=\tau R_{200},
\end{equation}
and adopt $\tau=3$ throughout this work. The BMO model retains an NFW-like behaviour in the inner part of the halo while providing a steeper decline in density at large radii. It therefore allows us to test the sensitivity of the weak-lensing mass and concentration measurements to the modelling of the outer cluster density distribution.
For both models, the free parameters are $M_{200}$ and $c_{200}$. 

In order to derive the lensing model quantities, we can compute the corresponding projected surface mass density as
\begin{equation}
\Sigma_{\rm model}(R)
=
2\int_R^\infty
\frac{\rho_{\rm model}(r)\,r}
{\sqrt{r^2-R^2}}
\,{\rm d}r \, ,
\end{equation}
from which we can derive the convergence for a given source and lens redshift.

The mean surface mass density enclosed within the projected radius $R$ is: 
\begin{equation}
\overline{\Sigma}_{\rm model}(<R)
=
\frac{2}{R^2}
\int_0^R
\Sigma_{\rm model}(R')\,R'\,{\rm d}R' \,,
\end{equation}
and thus we define the excess surface mass density as:
\begin{equation}
\Delta\Sigma_{\rm model}(R)
\equiv
\overline{\Sigma}_{\rm model}(<R)
-
\Sigma_{\rm model}(R)
=
\Sigma_{\rm crit},
\gamma_{{\rm t},{\rm model}}(R) , .
\end{equation}

Since our weak-lensing analysis is based on the reduced shear, the quantity entering the modelling is the reduced shear:
\begin{equation}
g_{{\rm t},{\rm model}}(R)
=
\frac{
\Delta\Sigma_{\rm model}(R)
}
{
\Sigma_{\rm crit}
-
\Sigma_{\rm model}(R)
} \, .
\end{equation}

For the NFW density profile, the projected surface mass density and the corresponding lensing quantities can be computed analytically following \citet{bartelmann96a}, 
which allows us to compute the lensing signal without numerically performing the line-of-sight projection of the three-dimensional density profile, which is needed for the BMO profile \footnote{On Apple M3, the calculation of the reduced excess surface mass density, in 32 logarithmically spaced bins from 0.1 $h^{-1}$Mpc to 2 $h^{-1}$Mpc of the BMO profile is $3 \times 10^3$ times slower than the analytical NFW, which takes 0.1 ms. This highlights that a full cluster shape analysis using a BMO profile will dramatically slow down the MCMC analysis.}

We explore the posterior probability distribution using a Monte Carlo Markov Chain (MCMC) analysis. Assuming Gaussian uncertainties, the logarithmic likelihood can be written as
\begin{equation}
\ln{\mathcal L}_{1D}
=
-\frac{1}{2}
\sum_i
\frac{
\left[
\Delta\Sigma_i-
\Delta\Sigma_{\rm model}(R_i|M_{200},c_{200})
\right]^2
}
{\sigma_{\Delta\Sigma,i}^2},
\end{equation}
up to an additive normalisation constant, where $\Delta\Sigma_i$ and $\sigma_{\Delta\Sigma_i}$ denote the measured profile and its uncertainty in the (i)-th radial bin, respectively. The same radial range, binning, likelihood, and priors (see first two lines of Table~\ref{priors}) are adopted for the NFW and BMO analyses, so that differences between their posterior distributions can be directly associated with the assumed radial mass model.

In Fig.~\ref{1D_prof}, the data points display the excess surface mass density measured for one realisation of the weak lensing simulations from the maps displayed in Fig.~\ref{Map_WSL}. The uncertainty on the binned data points has been estimated as the standard error of the mean, considering that the intrinsic galaxy shape noise is already contained in the catalogue values.

The blue and orange solid curves show the best-fit model from the MCMC to the data, considering NFW and BMO profiles, respectively. The reduced-shear correction is consistently accounted for in both the measured profiles and the corresponding model predictions.

\begin{table}[ht!]
\caption{\label{priors}Uniform priors adopted in our MCMC analyses}
\centering
\begin{tabular}{lc}
\hline\hline
Parameter & Priors range \\
\hline \\
$\log(M_{200}/(h^{-1}\mathrm{M_{\odot}}))$ & $[12,\,16]^*$ \\ 
$c_{200}$ & $[1,\,20]^*$ \\
$x_0$ & $[-0.1,\,0.1]\,h^{-1}$Mpc\\
$y_0$ & $[-0.1,\,0.1]\,h^{-1}$Mpc\\
$e$ & $[0,\,0.95]$ \\
$\phi$ & $[0,\,180]$ deg \\ \hline
\end{tabular}
\tablefoot{$^*$for the two-dimensional analysis, we perform an additional MCMC run
adopting Gaussian priors -- from the one-dimensional NFW fit -- on $\log(M_{200})$ and $c_{200}$, centred on their respective best-fitting values from the one-dimensional NFW analysis and with widths corresponding to the associated $1\,\sigma$ posterior uncertainties.}
\end{table}

Figure~\ref{fig_1DPost} compares the resulting $\log(M_{200})$ and $c_{200}$ posterior distributions for the NFW and BMO models of the three clusters and projections as presented in Fig.~\ref{1D_prof}. 
From the figure, we can notice that the assumed model can affect not only the best-fitting parameters but also their covariance. Mass and concentration are correlated, with a negative slope \citep{zhao09,giocoli12b}: changes in concentration modify the relative amplitude of the signal, which is balanced by a variation in mass.

   \begin{figure*}[ht!]
   \centering
   \includegraphics[width=\hsize]{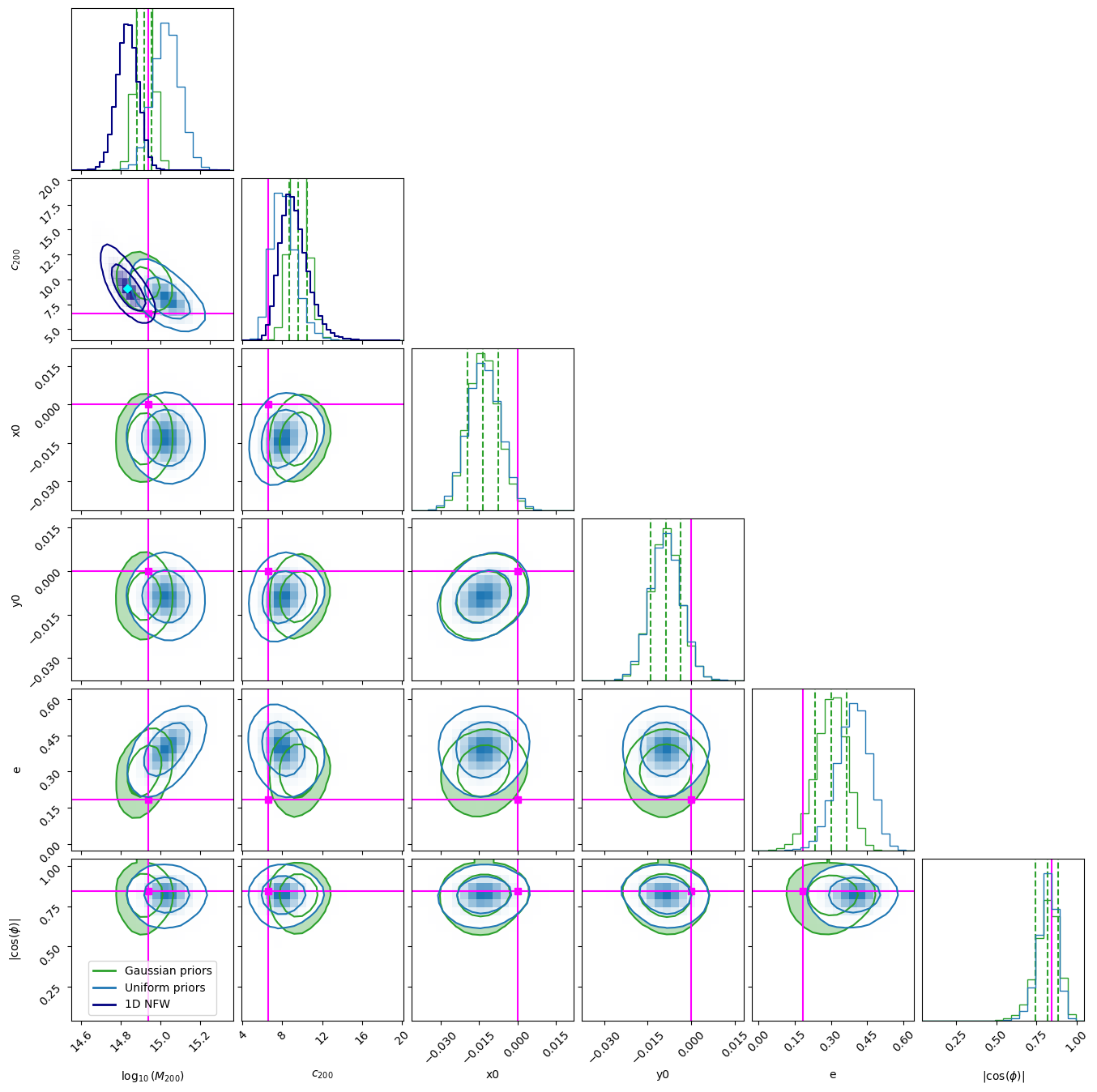}
      \caption{Posterior distributions with 1 and 2 $\sigma$ confidence regions, for one shear catalogue realisation, of the cluster morphological parameters: mass, concentration, and the centroid coordinates offset. Results are displayed for the projection $(X, Y)$ of cluster 129. The magenta lines indicate the reference cluster parameters. While the blue distributions consider uniform priors as in Table~\ref{priors}, the green histograms assume $\log(M_{200})$ and $c_{200}$ to be Gaussian centred on the posteriors of the one-dimensional fit, shown in dark blue colour in the top left panel.
         \label{2D_post}}
   \end{figure*}

\subsection{Two-dimensional NFW modelling}

For the two-dimensional weak-lensing analyses, the model parameters can be read as:
\begin{equation}
\boldsymbol{\Upsilon} =
\left[
\log_{10}(M_{200}),\,c_{200},\,x_0,\,y_0,\,e,\,\phi
\right],
\end{equation}
where $M_{200}$ and $c_{200}$ are the halo mass and concentration, $(x_0,y_0)$
is the centroid offset with respect to the projected coordinates of the maximum density peak, $e$ is the ellipticity, and $\phi$ is the position
angle of the major axis in the plane of the sky. 

Given a source position $(x,y)$, we rotated the coordinates around the
model centroid,
\begin{align}
x' &= (x-x_0)\cos\phi + (y-y_0)\sin\phi\,,\\
y' &= -(x-x_0)\sin\phi + (y-y_0)\cos\phi\,.
\end{align}
The equations above give us the possibility to write the elliptical radius as in the lens plane,
\begin{equation}
R_{\rm ell} =
\left[
x'^2 + \left(\frac{y'}{q}\right)^2
\right]^{1/2}.
\end{equation}
where $q=b/a=1-e$ is the projected minor-to-major axis ratio. The spherical NFW shear and convergence profiles are then evaluated at $R_{\rm ell}$, such that the radial dependence of the lensing signal follows elliptical rather than circular isocontours \citep{giocoli12a,giocoli12c,cui25}. The reduced tangential shear amplitude, $g_{\rm t}(R_{\rm ell})$, is projected onto its Cartesian components using the polar angle with respect to the fitted halo centre. 

This prescription should be regarded as a low-order elliptical approximation to the cluster lensing signal, rather than as a fully self-consistent triaxial or elliptical mass model \citep{adhikari15,fu26}. Its purpose is to provide a common parametric description of the dominant projected morphology, characterised by the centroid, position angle, and axis ratio, that can be applied uniformly to all cluster projections. This allows us to assess how well the main morphological properties encoded in the two-dimensional weak-lensing signal can be recovered and compared with those measured directly from the underlying convergence maps and from the tangential critical lines.

We assume a Gaussian likelihood in the two shear components, which can be read as:
\begin{equation}
\ln\mathcal{L}_{2D} =
-\frac{1}{2}
\sum_{j}
\frac{
\left[g_{1,j}-g_{1,\mathrm{model},j}(\boldsymbol{\Upsilon})\right]^2+
\left[g_{2,j}-g_{2,\mathrm{model},j}(\boldsymbol{\Upsilon})\right]^2
}
{\sigma_{\mathrm{e}}^2}\,.
\end{equation}
where $j$ denotes the background galaxies.  It is worth noting that we assume a diagonal covariance matrix, with independent intrinsic galaxy ellipticities and equal variance in both shear components. This approximation is appropriate for our controlled mock observations, where intrinsic alignments are not included, and the lensing signal is generated from fixed cluster projections. Correlations arising from intrinsic alignments, additional line-of-sight structures, and modelling uncertainties are therefore not included in either the simulations or the covariance.

To be consistent with the one-dimensional analysis, we exclude sources within 0.1 $h^{-1}$ Mpc. It is worth mentioning that, in the reduced-shear analysis, sources for which $|1-\kappa|\ll 1 $ are also excluded from the reduced-shear evaluation.

\begin{figure*}[ht!]
\centering
\includegraphics[width=1\hsize]{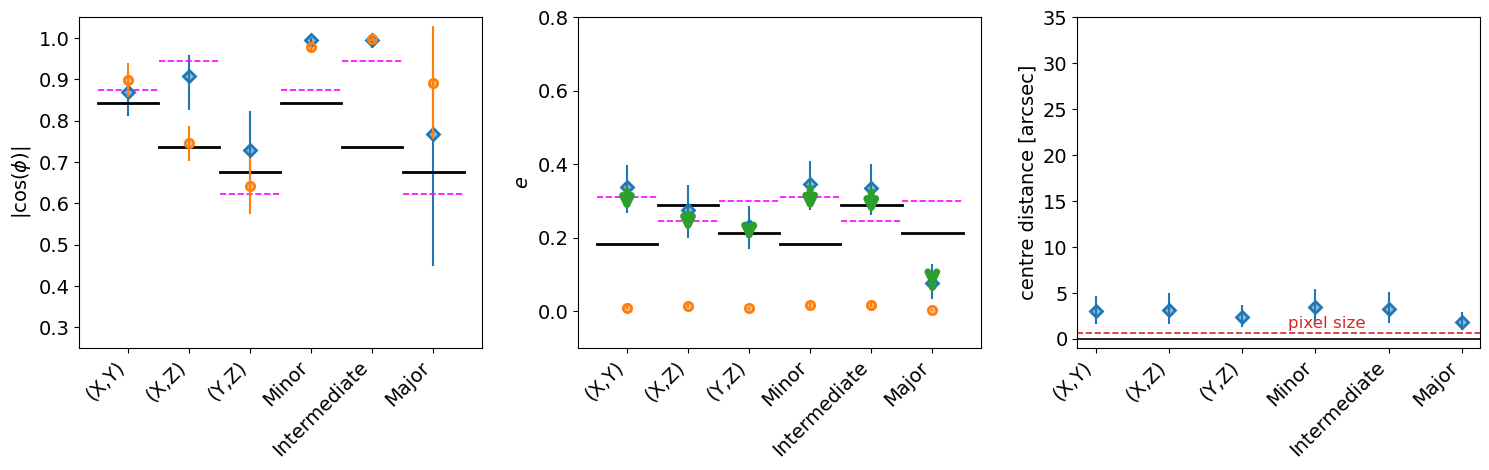}
\includegraphics[width=1\hsize]{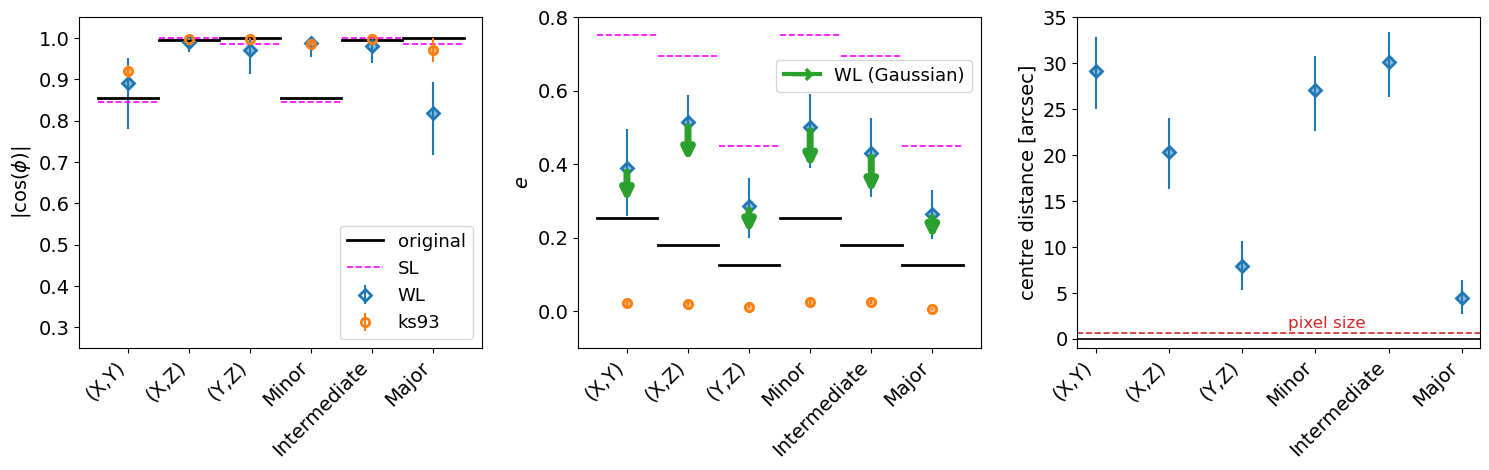}
\includegraphics[width=\hsize]{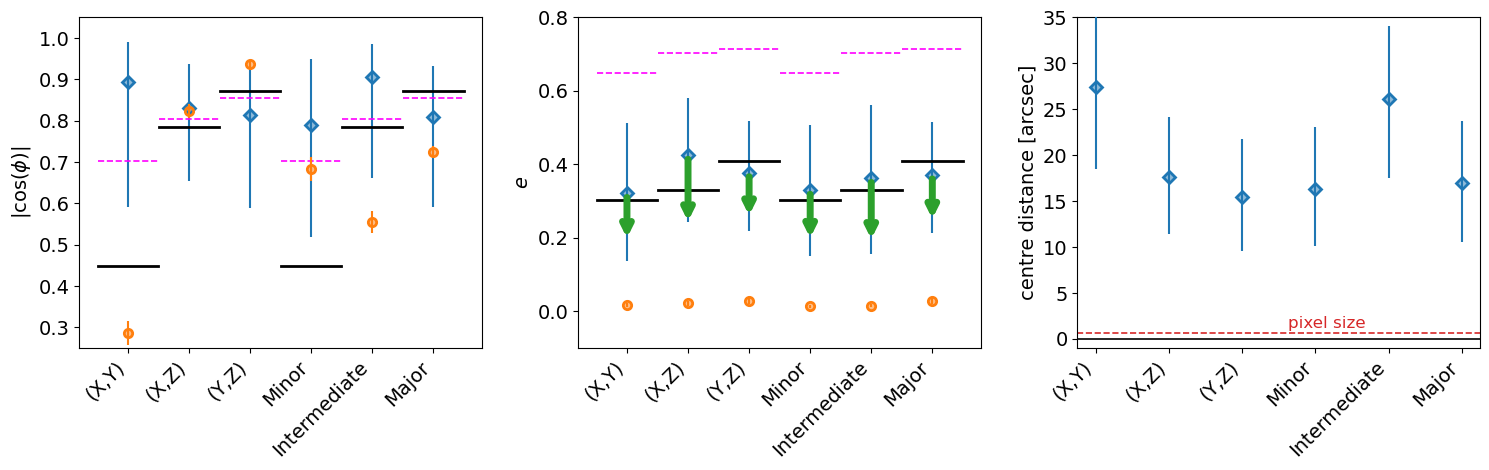}     
\caption{Panels from left to right, position angle, ellipticity and centroid shift for the various clusters, sorted from the relaxed to the most unrelaxed system, from top to bottom.  For each cluster, along the x-axis we display the running index of the different realisations modulo $128$. Solid black and dashed magenta lines indicate the average parameters computed from the WL and SL maps for each projection. The orange circles show the results from the KS93 reconstruction. Blue diamonds refer to the parameters computed from the two-dimensional NFW model using uniform priors, while the green arrow indicates the variation in ellipticity when assuming informative priors on mass and concentration as derived from the one-dimensional NFW fit.
\label{summary_plot}}
\end{figure*}

\section{Results}
\label{Results}

For each cluster, we perform the one- and two-dimensional weak-lensing analyses on 128 independent source and shape-noise realisations along six lines of sight, yielding 768 realisations per cluster. We present the recovered morphological parameters in blocks of 128 realisations, highlighting both the scatter within each projection and the systematic differences between viewing directions.

We independently reconstruct the convergence maps using the Kaiser \& Squires method \citep[][hereafter KS93]{kaiser93}. Since our catalogues contain reduced shear, we adopt the iterative inversion described by \citet{meneghetti_book}, stopping when the relative change between consecutive convergence maps falls below $10^{-4}$. We then measure the projected morphology from these maps, providing a non-parametric comparison with our two-dimensional modelling.

Figure~\ref{2D_post} shows the two-dimensional posterior distributions for halo 129 along the $(X,Y)$ projection, obtained with uniform (blue) and informative (green) priors on $\log(M_{200})$ and $c_{200}$ derived from the one-dimensional NFW fit. The corresponding one-dimensional posteriors are shown in dark blue in the top-left panel, while magenta markers indicate the reference two- and three-dimensional halo parameters. Examples for the other two clusters are presented in App.~\ref{app_2}.

In Fig.~\ref{summary_plot} we show the position angle, ellipticity and centroid shift offset recovered using the two-dimensional elliptical NFW model described in the previous section. From top to bottom, we sort clusters from the most relaxed to the most unrelaxed ones -- halo 129, 70 and 277, respectively. 

For each projection, the black horizontal lines indicate the morphological parameters measured directly from the true convergence maps and therefore provide our reference description of the projected cluster morphology on the scales probed by the weak-lensing analysis. The dashed magenta lines instead show the corresponding parameters inferred from the tangential critical lines. 
From the figure, we can see that for the relaxed cluster 129 (top row), the ellipticities (middle panel) inferred from the tangential critical lines and the larger-scale convergence distribution are relatively similar across most projections, suggesting a comparatively coherent projected morphology between the central and outer regions.
A different behaviour is observed in the two dynamically disturbed systems. The tangential critical lines indicate a more elliptical central mass distribution than that inferred from the convergence maps on larger cluster-centric scales. This difference indicates a stronger scale dependence of the projected morphology in dynamically complex systems, where mergers, asymmetries, and substructures can affect the inner and outer mass distributions differently. 

The blue diamonds show the average position angle, ellipticity and centre distance of our two-dimensional NFW model computed from the reduced shear catalogues. The values are computed by averaging over 128 shear catalogues and intrinsic shape-noise realisations for each cluster projection, while the error bars indicate their standard deviations. 
These results are obtained by adopting the uniform priors on the parameters listed in Table~\ref{priors}. Overall, the comparison with the reference values illustrates the ability of the two-dimensional weak-lensing analysis to recover the orientation and projected shape of the underlying mass distribution, while also highlighting the scatter introduced by the finite number density of background sources and their intrinsic ellipticities. We also notice that the error bars increase as we move from the relaxed to the unrelaxed system.

We repeat the two-dimensional analysis by incorporating the 
constraints on the radial mass distribution obtained from the one-dimensional 
weak-lensing analysis assuming the NFW profile. 
The resulting changes in the recovered ellipticity are indicated by the green arrows in the middle panels of Fig.~\ref{summary_plot}, while the other parameters remain unchanged. 

Compared to the analysis adopting uniform priors, introducing information on the radial mass distribution generally leads to a reduction of the inferred ellipticity. This behaviour highlights the covariance between the radial parameters of the halo model and its projected shape. When $M_{200}$ and $c_{200}$ are allowed to vary freely, part of the variation in the two-dimensional lensing signal can be accommodated through a combination of changes in the radial profile and in the projected ellipticity. Constraining the radial mass distribution with the one-dimensional analysis reduces this freedom and consequently modifies the ellipticity preferred by the two-dimensional model.

Finally, the orange circles show the mean morphological parameters measured from the KS93 convergence reconstructions. The reconstructed position angles generally follow those inferred from the two-dimensional weak-lensing modelling and the orientation of the underlying projected mass distribution. However, the ellipticity is systematically lower:  the KS93 reconstruction recovers rounder projected mass distributions. 
This behaviour is plausibly related to the combined effects of finite-source sampling, intrinsic shape noise, and the smoothing inherent in reconstructing a continuous convergence field from a discrete, noisy reduced-shear catalogue. These effects preferentially suppress small-scale and anisotropic features of the projected mass distribution, thereby diluting the recovered ellipticity while more effectively preserving the cluster's dominant orientation.

The right-hand panels of Fig.~\ref{summary_plot} show the offset between the recovered centre and the reference centre of the projected cluster mass distribution. The three systems display different behaviours. The relaxed cluster 129 exhibits comparatively small offsets, whereas larger displacements are found for the two dynamically disturbed clusters, particularly for halo 277. This behaviour is qualitatively consistent with their dynamical classification: in relaxed systems, the different definitions of the cluster centre tend to identify approximately the same region, whereas mergers and asymmetric mass distributions can produce significant differences between the density peak, centre of mass, and centre preferred by the weak-lensing model. The centroid offset, therefore, provides an additional morphological diagnostic that complements the ellipticity and position-angle measurements, which can be compared with the BCG position identified in real photometric data.

\section{Summary and Conclusions}
\label{sandc}
In this work, we have tested the ability of using lensing data to recover the two-dimensional 
structural parameters of individual galaxy clusters. Recovering that information is 
particularly important since its projected morphology reflects the triaxial structure, 
orientation, assembly history and dynamical state of their host dark matter haloes.

We selected three galaxy clusters at $z=$0.22 from The Three Hundred project as representative cases of a relaxed, a mildly disturbed and a strongly unrelaxed system. We have analysed each cluster along six different lines of sight: three random projections and three projections aligned with the minor, intermediate, and major axes of the three-dimensional mass tensor. This approach allowed us to investigate the impact of orientation on the recovered cluster morphology.

For each projection, we generated 128 reduced shear catalogues. 
We first modelled the one-dimensional excess surface mass density using NFW and BMO models, and subsequently exploited the two-dimensional shear catalogue to constrain the projected morphological cluster parameters: position angle, ellipticity, and centre. These measurements were compared with the morphology measured from the true convergence maps, with the convergence reconstructions obtained using the KS93 method, and with the shape of the tangential critical lines that trace the central strong-lensing region.

The main results of our analysis can be summarised as follows:
\begin{itemize}

\item the projected cluster morphology depends on the considered line-of-sight; a non-spherical halo can produce different projected ellipticities and orientation angles: projection effects matter when interpreting cluster lensing measurements;

\item the two-dimensional weak lensing signal contains information not only on the cluster mass and concentration but also on the orientation and ellipticities of individual systems. The recovered position angles follow quite well the orientation of the underlying projected mass distribution;

\item the radial and morphological parameters of the lens model are not 
independent. When the constraints on $M_{200}$ and $c_{200}$ obtained from 
the one-dimensional NFW analysis are used as informative priors in the 
two-dimensional modelling, the recovered ellipticity generally decreases. 
This demonstrates a covariance between the inferred radial mass distribution 
and projected shape and shows that assumptions or prior information on the 
radial profile can influence weak-lensing measurements of cluster 
ellipticity;

\item the projected morphology measured on a large scale and on the inner region depends on the dynamical state of the cluster. For the relaxed system, the inner and outer projected morphologies are relatively similar, whereas the two dynamically disturbed clusters show larger differences, with their central strong-lensing regions generally appearing more elliptical, in our controlled sample. This highlights how mergers and substructures tend to produce a complex multiscale projected morphology;

\item the KS93 projected mass reconstruction provides a complementary way to study cluster projected properties. However, while with this method the orientation angle is recovered quite well, the recovered convergence tends to appear rounder. It is worth mentioning that finite-source sampling, intrinsic shape noise, and the smoothing associated with the reconstruction can dilute the anisotropic component of the lensing signal while preserving its preferred orientation;

\item the recovered cluster centre, from two-dimensional mass modelling, provides complementary information about the complexity of the projected mass distribution. While relaxed systems exhibit a small offset with respect to the projected maximum density peak, unrelaxed and strongly disturbed clusters show a larger displacement. 

\end{itemize}

These results demonstrate that weak lensing can be exploited not only as a cluster mass estimator 
but also as a probe of the projected morphology of the matter distribution

While recent studies have demonstrated the potential of stacked
weak-lensing multipole analyses to constrain the average projected
ellipticity of galaxy cluster populations \citep{fu26},
our approach focuses on recovering the morphology of individual
clusters directly from their two-dimensional reduced-shear fields.
This provides complementary information on the projected orientation,
ellipticity, and centroid, and allows us to investigate their
dependence on the cluster dynamical state. Comparing these quantities with the morphology of the stellar and gas distributions, as well as with the orientation of the surrounding filamentary structures, could provide additional information on the cluster assembly history and three-dimensional structure.

On a more theoretical level, accurate measurements of cluster shapes may also provide a new avenue for investigating the nature of dark matter; halo shapes are sensitive to the dynamical properties of the dark-matter component and can differ between collisionless CDM and Self-Interacting DM \citep{despali25b,despali25,giocoli26}. In this regard, strong and weak lensing shape analyses could represent a complementary test for the nature of dark matter. 

We emphasise that the present analysis is based on a small, controlled
sample of three clusters selected to span different dynamical states; 
the trends identified here should not be interpreted as
statistical relations representative of the overall cluster population.
Extending this analysis to the full cluster population of The Three Hundred Project simulations will be
necessary to quantify the statistical significance of these trends and
their dependence on cluster mass, redshift, and dynamical state.

Such an extension will also provide the statistical calibration required
to correctly interpret projected weak-lensing morphology in forthcoming large
cluster samples from surveys such as Euclid\footnote{\url{https://www.euclid-ec.org}} \citep{euclidoverview}, Rubin\footnote{\url{https://rubinobservatory.org}}, and Roman\footnote{\url{https://science.nasa.gov/mission/roman-space-telescope}}.
Rather than compressing the lensing information exclusively into a mass estimate, these new data will offer the opportunity to use the projected shape and orientation of the matter distribution as additional cluster observables, combining them with strong-lensing, optical, X-ray, and SZ morphological information. 

\begin{acknowledgements}
MM  acknowledges financial support from the Italian Space Agency (ASI)
through contract ``Euclid - Phase E'' and from the INAF Grants ``The
Big-Data era of cluster lensing’’ and ``Probing Dark Matter and Galaxy
Formation in Galaxy Clusters through Strong Gravitational Lensing''.
LM acknowledges the financial contribution from the grant ASI n. 2024-10-HH.0 “Attività scientifiche per la missione Euclid – fase E”.  ER is supported by NASA EGIP Grants 80NSSC25K0006 and 80NSSC25K8009. 
\end{acknowledgements}

%\bibliography{globalbibs}
\bibliography{the300_cl_shapes}
\newpage

\begin{appendix}
\section{A view of the principal axis projections}
\label{app_1}

\begin{figure*}
    \centering
    \includegraphics[width=\hsize]{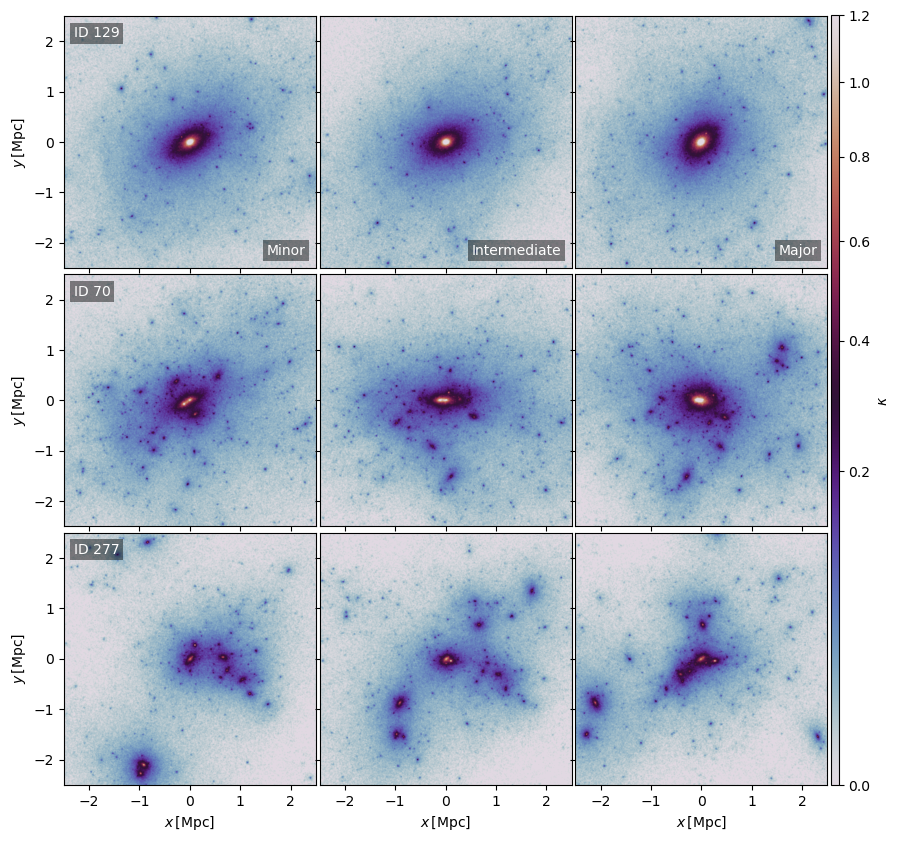}
    \label{Three_PR_Clusters}
    \caption{As Fig.~\ref{kappa_random} but for the projections along the minor, intermediate and major axes for the three clusters considered, from left to right, respectively. From top to bottom, we show clusters sorted from the most to the least relaxed. \label{kappa_part}}
\end{figure*}

In this appendix, we show in Fig.~\ref{kappa_part} the three projections along the minor, intermediate and major axes of the mass tensor ellipsoid. From the figure, we can see that the relaxed and mildly disturbed systems appear more elliptical when oriented along the minor axis than when oriented along the major axis. Less evident is the situation of the most disturbed cluster in the lower row, captured during a merging phase and composed of distinct substructures. 

\section{Additional posteriors for two-dimensional morphological parameters}
\label{app_2}

\begin{figure*}[ht!]
   \centering
   \includegraphics[width=\hsize]{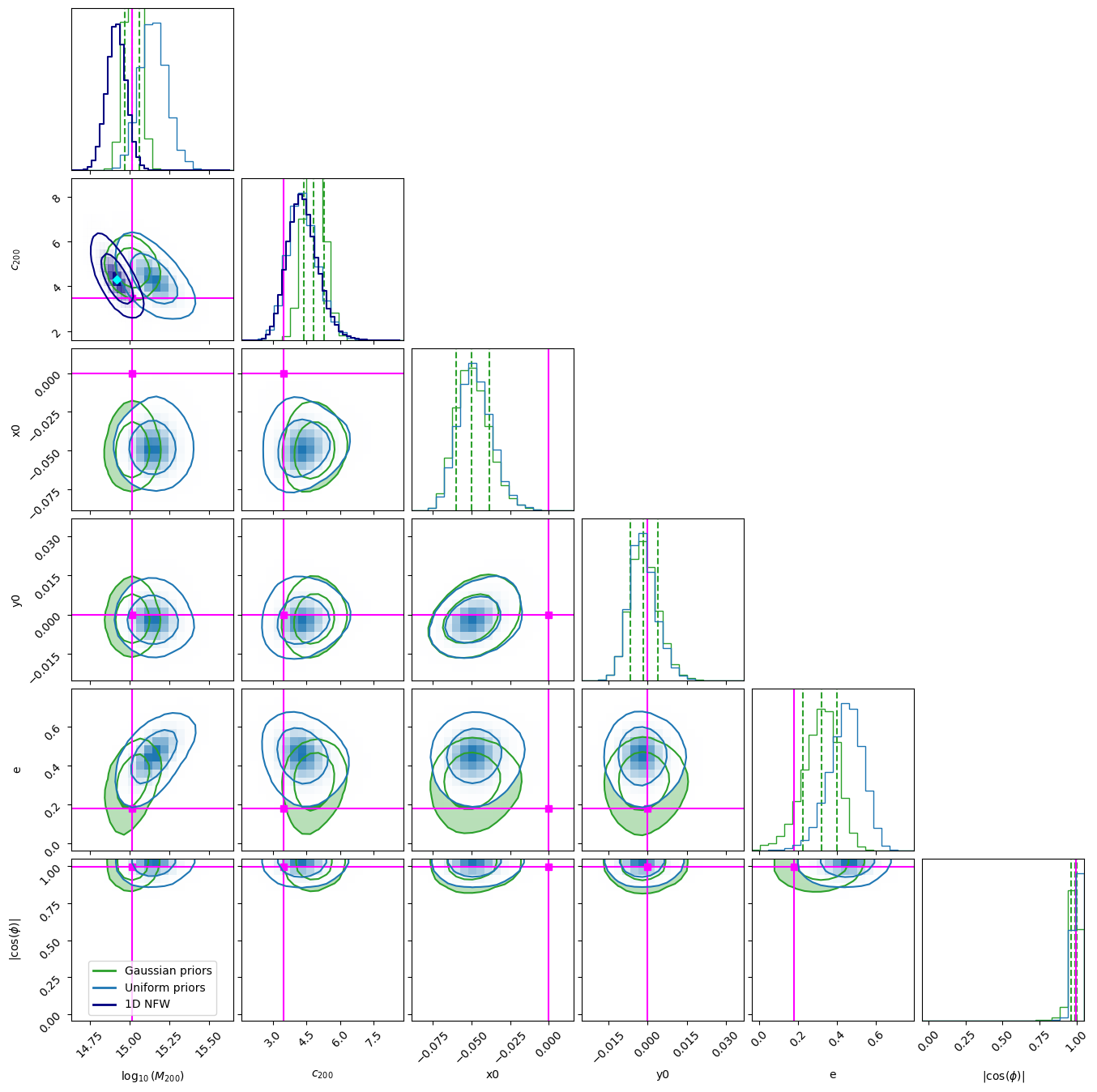}
      \caption{Posterior distributions for one shear catalogue realisation of cluster 70.
         \label{2D_post_bis}}
   \end{figure*}

\begin{figure*}[ht!]
   \centering
   \includegraphics[width=\hsize]{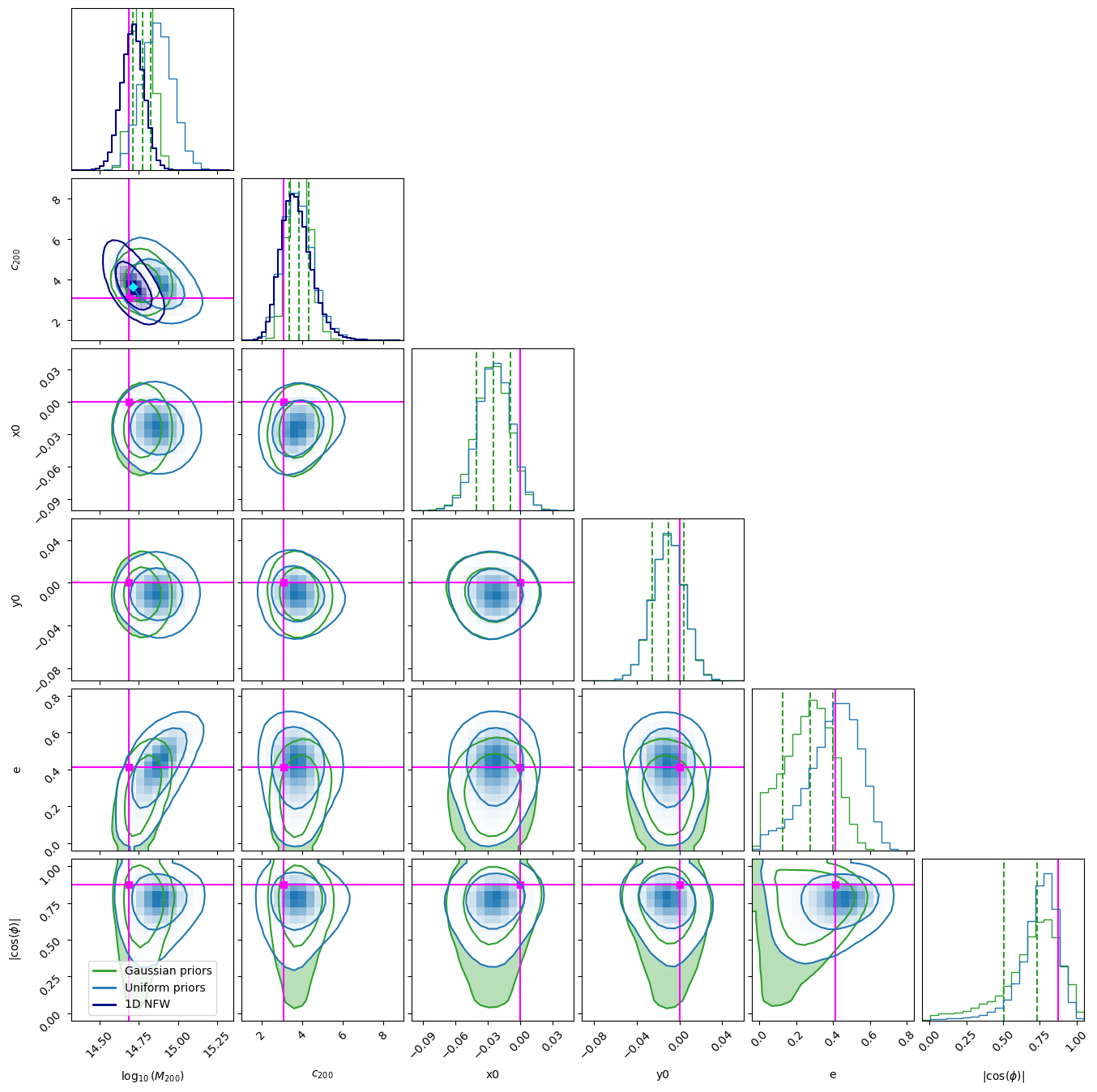}     
      \caption{As Fig.~\ref{2D_post_bis} but posterior distributions for one shear catalogue realisation of cluster 277.
         \label{2D_post_tris}}
   \end{figure*}

In this appendix, we display the posterior distributions of the two-dimensional parameters for one shear catalogue and an intrinsic noise realisation of the clusters 70 and 277, in Fig.~\ref{2D_post_bis} and Fig.~\ref{2D_post_tris}, respectively. As in Fig.~\ref{2D_post}, blue distributions refer to the case assuming uniform priors for all the parameters, while green distributions assume Gaussian priors for $\log(M_{200})$ and $c_{200}$ as derived from the one-dimensional analysis. Dark blue distributions, shown in the top-left panel, present the posteriors of the one-dimensional NFW fit, while the magenta lines and points correspond to the reference two- and three-dimensional cluster halo properties.

\section{Einstein radii table}
\label{app_3}

In this appendix, we summarise in Table~\ref{tab_Einstein} the measure of the Einstein Radii $\theta_E$ for each considered projection per cluster. Following \citet{meneghetti23}, the Einstein radius definition is computed from the area $A_{\rm crit}$ of the central critical lines as:
\begin{equation}
    \theta_E = \sqrt{\dfrac{A_{\rm crit}}{\pi}}\,,
\end{equation}
for sources at redshift $z_s=3$.

\begin{table}[ht!]
\caption{\label{tab_Einstein}Einstein Radii for each cluster projection in arcseconds.}
\centering
\begin{tabular}{lcccccc}
\hline\hline 
ID &$(X,Y)$ & $(X,Z)$&  $(Y,Z)$ & Minor & Intermediate & Major   \\
\hline \\
129 &  33.7 & 33.1 & 38.0 & 32.1 & 32.0 & 43.4 \\ 
$\;70$ & 15.4 & 19.2 & 26.6 & 16.8 & 16.7 & 33.9 \\
277 & 3.6 & 5.9 & 5.7 & 9.2 & 3.9 & 5.2 \\ 
\hline
\end{tabular}
\end{table}

\end{appendix}

\end{document}